# Infrared Characterisation of Jupiter's Equatorial Disturbance Cycle

Arrate Antuñano[1], Leigh N. Fletcher[1], Glenn S. Orton[2], Henrik Melin[1], John H. Rogers[3], Joseph Harrington[4], Padraig T. Donnelly[1], Naomi Rowe-Gurney[1] and James S. D. Blake[1]

[1] Department of Physics & Astronomy, University of Leicester, University Road, Leicester LE1 7RH, UK

[2] Jet Propulsion Laboratory, California Institute of Technology, 4800 Oak Grove Drive, Pasadena, CA 91109, USA

[3] British Astronomical Association, London, UK

[4] Planetary Sciences Group, Department of Physics, University of Central Florida, Orlando, FL 32816-2385, USA

**Key Points**

- We report a pattern of cloud clearance events that completely change the appearance of Jupiter's equatorial zone in the infrared.
- The equatorial disturbance forms in less than 1 month and dissipates over 4 months, with a typical lifetime of 12-18 months at 5 microns.
- A 6-8 or 13-14-year periodicity is found. We predict a new equatorial zone disturbance at 5 μm to occur in 2019-21.

**Abstract**

We use an infrared dataset captured between 1984 and 2017 using several instruments and observatories to report five rare equatorial disturbances that completely altered the appearance of Jupiter's Equatorial Zone (EZ): the clearance of tropospheric clouds revealed a new 5-µm-bright band encircling the planet at the equator, accompanied by large 5-µm-bright filaments. Three events were observed in ground-based images in 1973, 1979 and 1992. We report and characterize for the first time the entire evolution of two new episodes of this unusual EZ state that presented their maximum 5-µm-brightness in December 1999 and February 2007, coinciding with a brown coloration south of the equator and with large bluish filaments and white plumes in the northern EZ at visible wavelengths. We characterize their typical infrared-bright lifetimes of 12-18 months, with possible periodicities of 6-8 or 13-14 years. We predict that a full-scale equatorial disturbance could occur in 2019-21.

1. Introduction

Jupiter's Equatorial Zone (EZ) displays unique atmospheric phenomena that are not observed at other latitudes, the most remarkable of which is a compact and quasi-periodically spaced pattern at the northern region of the zone at ~7°N (EZ(N)) planetocentric latitude (all latitudes in this study are planetocentric), formed by white plumes (potentially freshly-condensed $NH_3$ ice) extending equatorward, and compact dark blueish-grey features, known as 5-µm hotspots (Keay et al., 1973; Terrile and Westphal, 1977; Beebe et al., 1989; Rogers, 1995;

Ortiz et al., 1998). These hotspots have been found to be regions of depleted volatiles and aerosols (Wong et al., 2004; Fletcher et al., 2016; de Pater et al., 2016), moving ~70 m/s slower than their surroundings (Beebe et al., 1996; Arregi et al., 2006; García-Melendo et al., 2011; Asay-Davis et al., 2011; Choi et al., 2013), possibly as a result of strong subsidence in the presence of a Rossby wave trapped at the southern edge of the North Equatorial Belt (NEBs) near 6°-7°N (Allison, 1990; Friedson, 2005).

Interestingly, the southern edge of the EZ (EZ(S)) displays a very different cloud morphology of small periodic chevron-like features at ~ 7.5°S that change in short time periods (Simon-Miller et al., 2012), and which have been reported to be inertia-gravity waves (Allison, 1990) or Rossby waves (García-Melendo et al., 2011; Simon- Miller et al., 2012). At some epochs, a large feature known as the South Equatorial Disturbance (SED) is also present, extending from south the equator to the EZ(S) (Rogers, 1995; Sánchez-Lavega and Gómez, 1996; Simon-Miller et al., 2012; Rogers et al., 2013), resembling the EZ(N) plumes and suggesting, together with the motion of the chevrons-like features, a presence of an EZ(S) Rossby wave mirroring that in the north (Simon-Miller et al., 2012; Rogers et al., 2013). All this cloud morphology, however, is not usually observable in the deeper cloud-forming region at 1-4 bar, where 5-μm data usually reveal the quiescent EZ to be covered in thick cloud cover away from the hotspots (Westphal, 1969; Orton et al., 1998).

Since at least the late 19[th] century, Jupiter's EZ often displayed a planetary-scale transient change at visible wavelengths seen as an ochre/brownish coloration (Peek, 1958; Beebe et al, 1989;

Rogers, 1995), the last event being observed in 2012-2013 (Fletcher, 2017). These coloration events are often associated with highly variable dark and narrow filaments (festoons) emerging from the visibly dark 5-μm hotspots at the EZ(N), and accompanied by the EZ(N) white plumes. This variability of the equatorial zone seemed to be periodic at visible wavelengths, with an 8-year periodicity at the upper-tropospheric hazes (at 410 nm) and a 12-year periodicity in the ultraviolet (Simon-Miller and Gierasch, 2010).

Ground-based images at 5 μm from June-November 1973, March 1979 and January-April 1992 (see Table 1 for references) showed an unusual morphology in the Equatorial zone with a bright band encircling the planet at the equator, accompanied by large bright festoons emanating from the hotspots at ~7° N. In this paper, we report two new disturbances at the EZ that were observed in ground-based 5-μm images captured at NASA's Infrared Telescope Facility (IRTF) between September 1999 and August 2000, and between April 2006 and September 2007. The large temporal coverage of ground-based 5-μm infrared observations allows us to analyse the brightness variability of the EZ over almost 3 Jovian years from June 1984 to August 2017, characterizing for the first time the formation and evolution of the EZ disturbances at 5 μm and comparing them with the observed morphology of the cloud tops, in order to improve our understanding of the relationship between the cloud morphology, coloration, and the large-scale upwelling at Jupiter's equator (Achterberg et al., 2006; Fletcher et al., 2016; de Pater et al., 2016; Li et al., 2017).

## 2. Observations

In this study, we have used images captured over almost 3 Jovian years, between June 1984 and August 2017, by 8 different ground-based instruments (see table 1) at infrared wavelengths between 4.76 µm and 5.18 µm. The 3-m Infrared Telescope Facility (IRTF) provides diffraction limited spatial resolutions of ~0.4" over the 4.76-5.18 µm range, compared to the 0.16" for 5.1 µm of the 8-m Very Large Telescope (VLT). However, data were acquired under average seeing conditions of ~1", which corresponds to a spatial resolution on Jupiter's equator of ~2850 km (2.3º longitude) at opposition, or ~3800 km (3.1º longitude) at quadrature, sufficient to perform detailed analysis of the variability of the cloud morphology and dynamics at this region.

All the ground-based instruments used in this study, except the TEXES instrument, acquire the data at several positions in a single observation, by nodding the telescope between different beams (Fletcher et al., 2009), enabling us to detect Jupiter's emission on top of the sky and instrumental background. In the case of the TEXES instrument, the data are obtained by stepping the spectrograph slit across the disk of Jupiter, with the first and last exposures providing a measure of the telluric background (Fletcher et al., 2016). See the supporting information for the data reduction method.

Table 1. A summary of the date of observations, instruments, number of images captured with the different instruments and wavelengths and the image scale of each instrument used in this study and the dates and data

showing the 5 EZ disturbance events. Note that all the instruments used, except VISIR that is located in Paranal (Chile), are mounted at the Infrared Telescope Facility (IRTF) on Maunakea, Hawaii.

## DATA USED IN THIS STUDY

| Date | Instrument | Number of images | Wavelength (μm) | Pixel Scale ("/pixel) | Configuration | References |
|---|---|---|---|---|---|---|
| Jun 1984 – Feb 1991 | BOLO-1 | 49 | 4.80-4.90 | - | Raster-scanned | Orton et al., 1994 |
| Jan 1992 – Apr 1992 | ProtoCam | 229 | 4.85 | 0.14-0.35 | Mosaicked | Harrington et al., 1996a,b |
| Apr 1994 – Sept 1994 | NSFCam | 6 | 4.85 | 0.15 | Full-frame | Ortiz et al., 1998 |
| Apr 1995 – Feb 2004 | NSFCam | 646 | 4.78 | 0.15 | Full-frame | Ortiz et al., 1998 |
| Jan 2005 – Apr 2006 | MIRSI | 17 | 4.9 | 0.27 | Full-frame | Fletcher et al., 2009 |
| Apr 2006 – Oct 2008 | NSFCam2 | 79 | 4.78 | 0.04 | Full-frame | Fletcher et al., 2010 |
| Jul 2009 – May 2015 | SpeX | 70 | 4.76 | 0.12 | Full-frame | - |
| Nov 2015 – Aug 2017 | SpeX | 133 | 5.1 | 0.12 | Full-frame | - |
| Dec 2014 – Jul 2017 | TEXES | 25 | 4.68 | 0.14 | Scanned | Fletcher et al., 2016 |
| Mar 2015 – Jan 2017 | TEXES | 47 | 5.18 | 0.14 | Scanned | Fletcher et al., 2016 |
| Dec 2016 – July 2017 | VISIR | 57 | 5.0 | 0.04 | Full-frame | Fletcher et al., 2017 |

## DATA SHOWING THE EZ DISTURBANCE EVENTS

| Date | Instrument/Telescope | Reference |
|---|---|---|
| [1] June - Nov 1973 | Catalina Observing Station, Hale | Armstrong et al., 1976, Westphal et al., 1974, Terrile and Beebe, 1979 |
| [2] March 1979 | Hale & IRTF | Terrile et al., 1979 |
| [3] Jan –Apr 1992 | ProtoCam | Harrington et al., 1996a,b; This Work |
| [4] Sept 1999 – Aug 2000 | NSFCam | This work |
| [5] Apr 2006 – Sept 2007 | NSFCam2 | This work |

3. Results

Images at 5 μm captured by both the NSFCam instrument between September 1999 and August 2000, and by the NSFCam2 instrument between April 2006 and September 2007, showed the formation and evolution of an unusual perturbation that arose rapidly in the EZ, leaving a bright 5-μm band encircling the planet between 1°S and 5°S, accompanied by large bright festoons emanating from the hotspots at ~7° N (see Figure 1G). RGB images of Jupiter, captured by amateurs during the same period (Hueso, 2018) also clearly show an upper-tropospheric perturbation at cloud level: (i) the EZ(N) displays a light grey/brownish colour with a number of white plumes bounded by dark grey/blueish festoons, expanding equatorward, and (ii) the EZ(S) presents a darker brown-and-grey band at 1-5° S (Figure 1H and Figure 1I). In March 2007, two similar festoon-like features also extended equatorward from the northern edge of the South Equatorial Belt (SEBn), related to the variable SED that arises irregularly at ~7.5°S (Simon-Miller et al., 2012; Rogers et al., 2013). So far, several such coloration events have been observed at visible wavelengths, however, it was not possible to determine how these colour changes corresponded to variation in the deep 1-4 bar cloud opacity until the advent of the 5-μm observations.

Surveying the literature, we found that three EZ disturbance episodes at 5 μm were observed before this work (but were not discussed at all), in June-November 1973, March 1979 and January-April 1992 (see Figure 1 and Table 1), indicating that a disturbance at the EZ at 5 μm could develop every 6-13 years, similar to the ~8-year periodicity observed in this region at the upper troposphere at 410 nm and the ~12-year periodicity found in the UV (Simon-Miller and Gierasch, 2010).

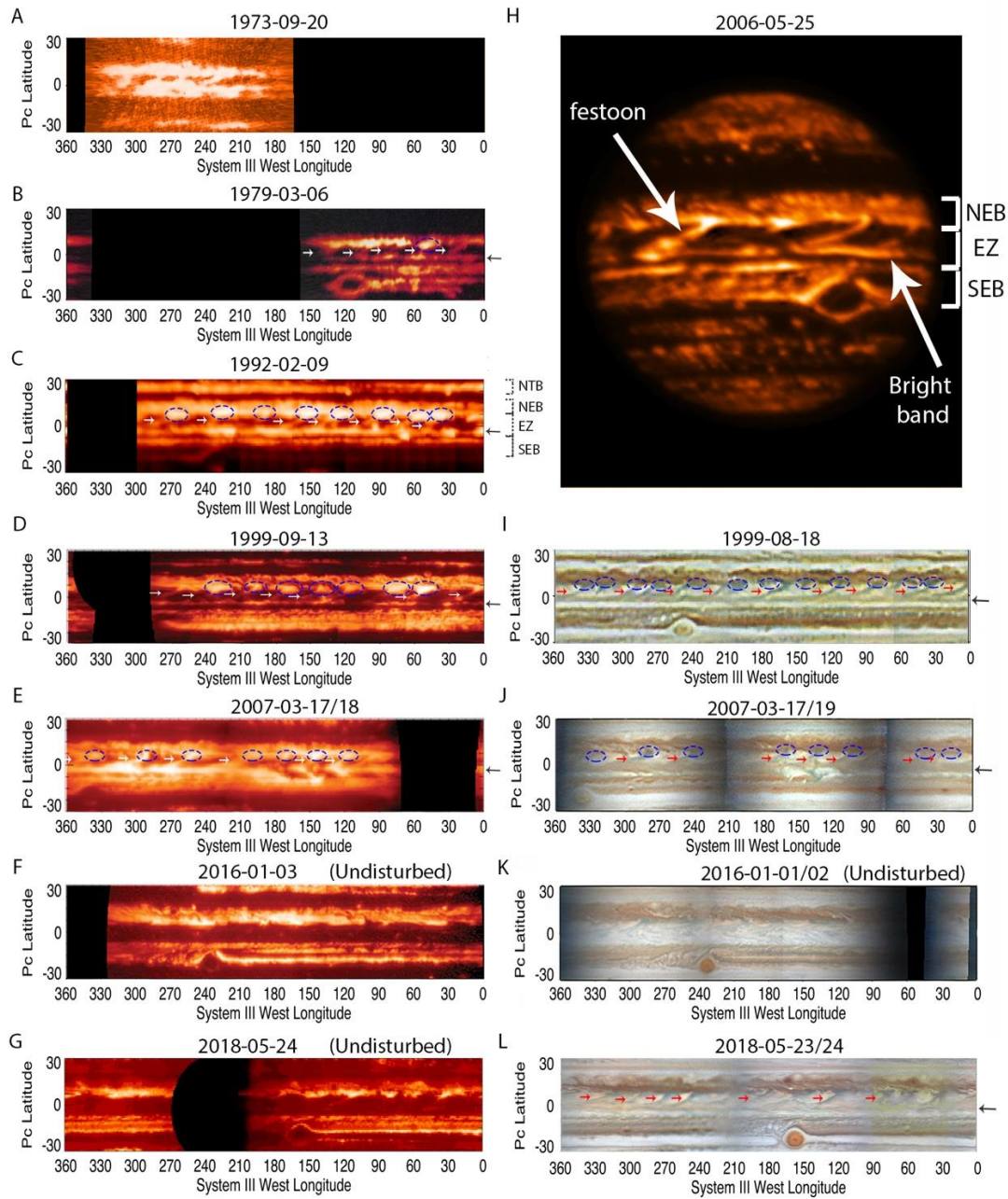

Figure 1. Cylindrical projections of images of Jupiter between ±30° latitude showing the five observed EZ disturbances at 5 μm (A-E) and their appearance at visible wavelengths (I and J), compared to more typical conditions of the EZ in 2016 at 5 μm (F) and at visible wavelengths (K). This is compared to Jupiter's appearance in the current apparition (May 2018) at 5 μm (G) and visible wavelengths (L), where a strong coloration is observed accompanied by dark festoons, but no EZ disturbance is yet visible. Panel H shows an image of Jupiter during the EZ disturbance. Blue dashed ellipses indicate the hotspots. The festoons are indicated by white arrows in A-E and by red arrows in I, J and L. The black arrows indicate the bright band at 5 μm (brownish at visible wavelengths) at 1°S-5°S. Panel A and B are built using Figure 1a in Westphal et al. (1974) and Figure 1 in Terrile et al. (1979), respectively. See Table 1 for data sources of the rest of the cylindrical maps at 5 μm.

Figure 2 displays the normalised and longitudinally-averaged brightness (see the online supporting information) of the EZ for different dates between June 1984 and August 2017. Analysing the brightness variability over the epochs studied, we find a distinct increase of the average brightness within the EZ during 1992 (blue), 1999-2000 (green) and 2006-2007 (orange) (Figure 2a and Figure2b), irrespective of the choice of how the data are scaled and normalised. These increases are not centred at the equator, instead covering latitudes between ~3°N and ~7°S at all epochs, displaying their maximum average brightness at 2°S - 3°S. These increases, which are clear signals of the presence of an EZ disturbance within and below the ammonia condensation level near 700 mbar, exhibit lifetimes of at least (i) 4 months, (ii) ~ 12 months and (iii) ~16 months, being observable at 5 µm between (i) January and April 1992, (ii) August 1999 and August 2000, and (iii) April 2006 and September 2007, respectively (see Figure 2c).

The lack of 5-µm data between May 1991 and January 1992 does not allow us to constrain the starting date of the EZ disturbance of 1992, however, we know that it must have started after May 1991. The short (4-month) temporal coverage of the 1992 event does not allow us to analyse the entire evolution of the disturbance, and therefore, it will be omitted in the comparison and description of the EZ disturbances.

During the 1999-2000 and 2006-2007 bright EZ events, the brightness at 5 µm increased very rapidly, reaching their maximum average brightness on 31 December 1999 and on 25 February 2007, around 6 and 10 months after the first bright spots appeared on the EZ, respectively. The average brightness is found to be higher at latitudes between 1°S and 5°S, where the 5-µm bright band (grey/brownish band in the visible) encircling the planet is observed, while the region between the equator and 3°N, where the festoons are located, exhibits around half of the maximum brightness. This difference is likely due to the horizontal averaging performed, as the festoons are observed to be as bright as the band between 1°S and 5°S. The peak observed in late 1989 and early 1990 at around 2° N in Figure 2c, corresponds to an unusual narrowing and southward shifting of the North Equatorial Belt (NEB) and it is not related to the EZ disturbances described in this work.

Unfortunately, the relative brightness of the three disturbance events might not be significant, as they depend on the chosen scaling region used to normalise the uncalibrated data (the South Tropical Zone being the optimum choice as it displays the least variability, see the online supporting information) and therefore, one should be cautious when comparing the intensity from event to event. However, the temporal variations are robust irrespective of the chosen scaling region.

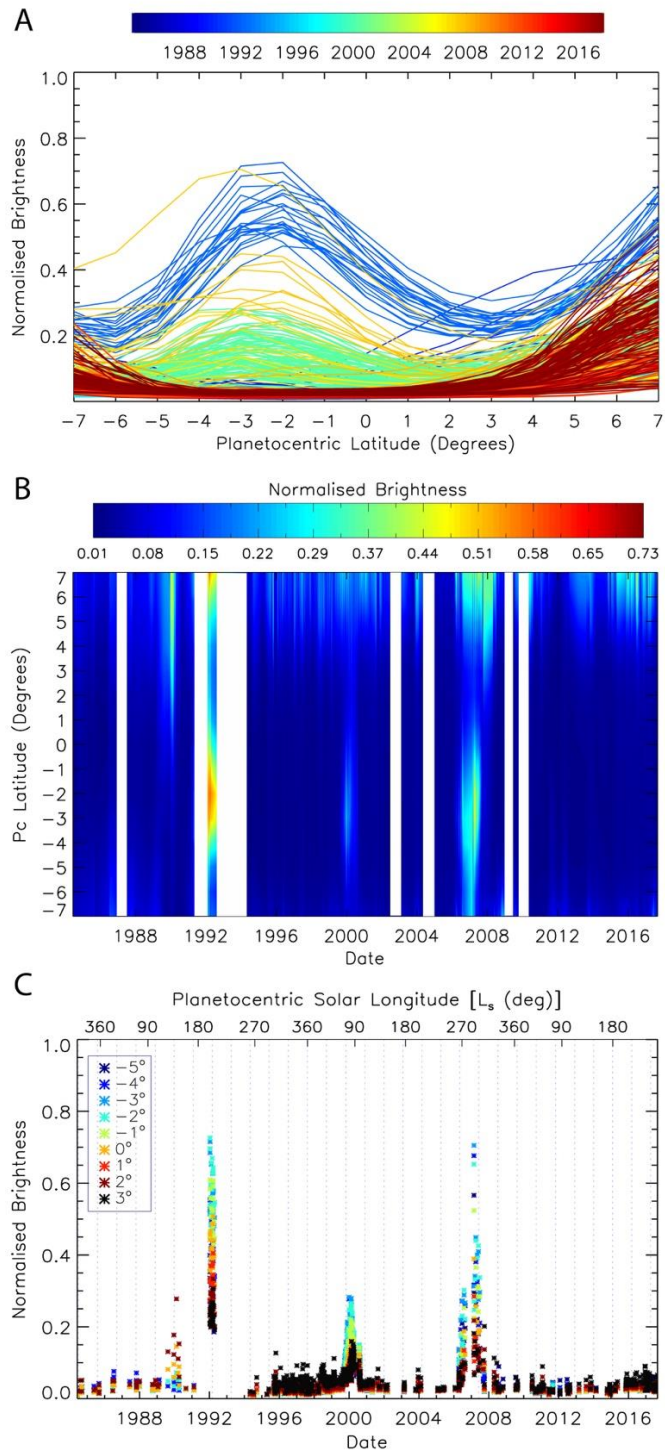

Figure 2. Temporal variability of (A) the normalised zonally averaged brightness at 5 μm between 7°N and 7°S over almost three Jovian years between June 1984 and August 2017, (B) contour map of the brightness against date, and (C) the normalised zonally averaged brightness as a function of time for the region between 3°N and 5°S. The blue dotted lines in (C) correspond to Jupiter opposition dates. White regions in (B) represent the gaps in the sampling larger or equal to 9 months.

The evolution of the 1999/2000 and 2006/07 EZ disturbances are shown in Figure 3 (see Figure S2 in the online supporting information for the evolution of the EZ disturbance between January and April 1992). The EZ disturbance at 5 µm from 1999 to 2000 (prior to Cassini's flyby in December 2000), started in August 1999, when a bright region extending over 40° of longitude appeared between 2°S and 4°S, accompanied by various hotspots at ~7°N with narrow and slightly bright tails (precursors of the longer festoons) with lengths of around 14° ± 1° longitude. The disturbance grew rapidly with longitude: within 3 weeks most of the longitudes between 2°S and 4°S appeared bright, and the hotspots had already developed large festoons that reached as far south as the bright band south of the equator. Two months after the first bright spots were observed at 5 µm, the NEBs hotspots presented festoons with longitudinal sizes between 20° ± 1° and 30° ± 1° that connected the NEBs and the 5-µm-bright band, forming a wave-like longitudinal pattern. During the following 3 months, the EZ disturbance kept growing, developing brighter festoons and a wider and brighter band that, by the time of the maximum brightness at the end of December 1999, covered the latitudes between 1°S and 5°S.

During this EZ disturbance 11 different hotspots (and festoons) were observed at 5 µm, moving with zonal velocities between 101.2 ± 0.3 m/s and 104.1 ± 0.3 m/s with respect to System-III rotation period (Archinal et al., 2011) (see Figure S1 in the online supporting information), which is around 50 – 70 m/s slower than their surroundings (García-Melendo et al., 2011), in agreement with the number of festoons and drift rate present at visible wavelengths during this period (Arregi et al., 2006; Rogers et al., 2013).

By April 2000, the EZ disturbance started to dissipate, and within 4 months Jupiter's equator had returned to its usual, quiescent appearance. By the time of Cassini's flyby in December 2000, there was no evidence of residual equatorial 5-μm brightness in VIMS observations (Giles et al., 2015; Sromovsky and Fry, 2010).

The EZ disturbance at 5 μm from 2006-2007 (and overlapping with the New Horizons flyby in February 2007, Baines et al., 2007), underwent a similar evolution. The first 5-μm images showing the activity of this new episode were captured on 12 April 2006. By this time, the disturbance was already quite developed, displaying 11 NEBs hotspots and various bright regions between 2°S and 4°S. The lack of 5-μm data between January and April 2006 does not allow us to determine the exact starting date of this disturbance. In May 2006, the festoons were already observable, connecting the NEB and the bright band south of the equator, showing an identical wave-like structure to that of 29 October 1999. Over the following 9 months, the bright band encircling the planet became wider and covered latitudes between 1°S and 5°S, displaying a non-uniformly distributed brightness.

Images captured in March 2007 showed a very singular appearance of the disturbance, not observed at the 1999-2000 event, displaying two bright festoons arising from the SEBn at ~7°S and connecting with the bright band at 1°-5° south, mirroring the festoons at the EZ(N). These southerly festoons, which are associated to the South Equatorial Disturbance (Simon-Miller et al., 2012; Rogers et al., 2013), were not present in images captured in February 2007 and evolved rapidly over the following weeks to reach their maximum length on 17 March 2007, with longitudinal sizes around 21° ± 1°. Images acquired only 3 weeks later no longer showed these southerly festoons. After that, the EZ displayed again the previous wave configuration of bright festoons and the band south of the equator until at least June 2007.

During all these months, 11 festoons were again observed moving with a mean zonal velocity of 104.9 ± 0.7 m/s, similar to previous measurements of the velocity of the hotspots at visible wavelengths (Arregi et al., 2006; Hueso et al., 2017). By the beginning of September 2007, the festoons started to dissipate and the 5-μm bright band began to darken. A month later, the EZ disturbance at 5 μm was no longer present. The dissipation timescale for the EZ disturbance at 5 μm from 2006 to 2007 was less than 4 months, although the absence of data between June and September 2007 does not allow us to obtain a more precise dissipation timescale.

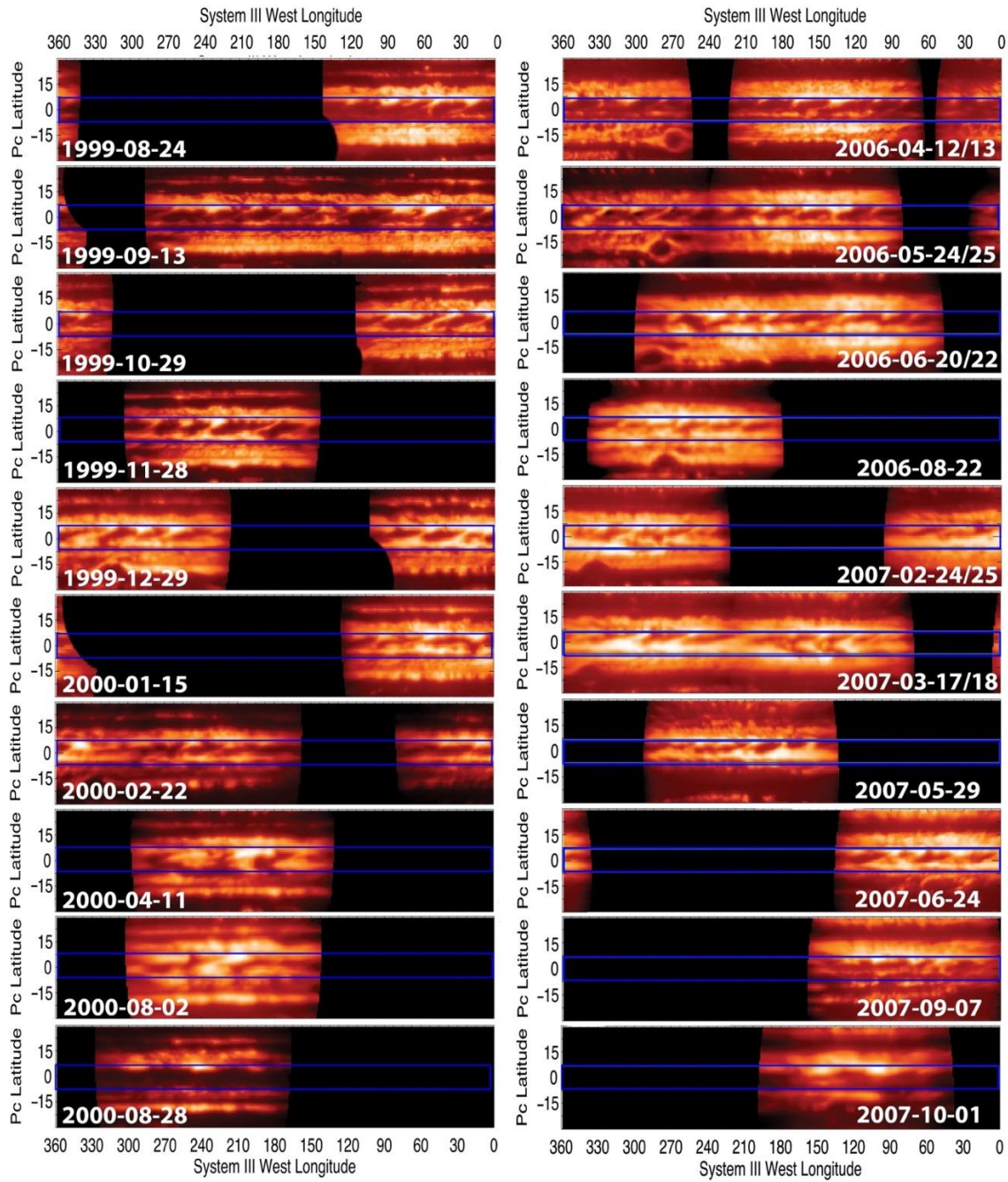

Figure 3. Time series of cylindrical projections of 5-μm images of Jupiter between 30°N and 30°S, showing the evolution of the two Equatorial Zone disturbances from 1999-2000 (left) and 2006-2007 (right). The EZ is highlighted in blue.

The observed lifetimes of the two EZ disturbances at 5 µm described above are found to be shorter than the lifetimes of the coloration events observed at the cloud tops, which usually last around 3 years (Rogers, 1995). In the case of the 2006-2007 EZ disturbance at 5 µm, RGB images captured by amateurs show that the EZ started to display an ochre coloration ~18 months before bright regions developed at 5 µm, while the presence of large grey/bluish festoons and white fans at cloud tops started to be observable around March 2006, some 5 months before the first signs of the EZ disturbance at 5 µm appeared. This indicates that the aerosol changes begin at high altitudes in the upper troposphere, and take many months to propagate down to remove the aerosols in the 1-4 bar region sensed at 5 µm. By the time of maximum brightness at 5 µm, RGB images showed large white plumes at the EZ(N) forming a wave-like pattern and a dark brown EZ(S).

4. **Discussion and Conclusions**

In this paper, we have identified five extreme episodes of cloud-clearing and 5-µm brightening at the equator since the 1970s, which completely changed the appearance of Jupiter's equatorial zone, and characterised in detail the 1990-2000 and 2006-07 events, tracking the evolution of a cloud-free band south of the equator and the narrow and bright festoons emerging from the hotspots. These disturbances lasted for 12-18 months at 5 µm, with formation and dissipation timescales of less than a month and less than 4 months, respectively.

The differences of the cloud morphology observed in the northern and southern EZ during the EZ disturbances could be related to the asymmetry of the ammonia found in the EZ, where the ammonia is elevated over the EZ(N) compared to the EZ(S) (Achterberg et al., 2006; Fletcher et al., 2016; de Pater et al., 2016; Li et al., 2017) reinforcing the idea that the plumes, festoons and hotspots are related to an NEBs Rossby wave, with ammonia uplift feeding the cloudy and 5-µm-dark plumes and subsidence depleting the hotspots and festoons of aerosols and gases (Fletcher et al., 2016; de Pater et al., 2016; Cosentino et al., 2017). During disturbances, the wave pattern is observed to extend southwards through the equator, together with a disturbed EZ(S) between 1°S and 5°S, both at cloud level and at the 1-4 bar level. This raises the possibility that this wave pattern is always present across the whole equatorial zone (e.g., as far south as 7°S), but that it is usually hidden by the thick ammonia clouds at higher altitude until an EZ disturbance event occurs. Alternatively, the wave pattern may intensify and extend southwards during these disturbance epochs.

Establishing a causal connection between the visible coloration episodes and the 5-µm disturbance is challenging – cloud thinning may be responsible for the early reddening/darkening at p<1 bar (still 5-µm dark), which sometimes (1973, 1979, 1992, 1999, 2006) but not always (1986-87, 2012-13) can lead to the full disturbance at 5 µm (i.e., removing clouds at p~1-4 bar). This is in agreement with a decrease of the albedo observed at 1.58 µm (cloud clearance of the ammonia cloud tops) during the 2007 EZ disturbance (see Figure S3 in the online supporting information), the decrease in the brightness observed at 410 nm (i.e., a removal of the tropospheric hazes) in 1999 and 2007 (Simon-Miller and Gierasch, 2010) and the decrease in the reflectivity of the tropospheric clouds observed by Lii et al (2010) in 1999-2000 and 2007. Interestingly, a more complex temporal variability of the brightness of the stratospheric hazes is observed at 2.12 μm, but is not clearly related to the EZ disturbances at 5 μm, suggesting that the EZ disturbances are confined to the cloud deck (see Figure S4 in the online supporting information). Future radiative transfer studies will be essential to understand possible changes in the clouds, hazes, temperature and ammonia variability associated with these events.

After examining five documented EZ disturbances, we conclude that they are periodic, with typical intervals of 6-8 years or 13-14 years, with "missing events" in 1986-87 and 2012-13 providing a source of confusion. These time intervals are also found in the upper troposphere at 410 nm and in ultraviolet wavelengths (~8-year and ~12-year periodicity, respectively (Simon-Miller and Gierasch, 2010)) and in changes on Jupiter's zonal wind profile at the equatorial latitudes (Tollefson et al., 2017), where the zonal winds seem to be faster than usual at the equator during or just after the EZ disturbance is observed at 5 μm. This latter observation supports the idea of equatorial cloud clearance during these events to permit the tracking of features located deeper in the atmosphere where zonal velocities are greater (1-4 bar, e.g. festoons and features inside the 5-μm bright band at the EZ(S)). The quasi-periodic timescale for these EZ disturbances could hint at some moist convective process, with the periodic release of accumulated potential energy, maybe even triggered by changes to the Rossby wave pattern. Future numerical simulations will be essential to understand these rare events.

Two of the five EZ disturbance events at 5 μm (the 1992 and 2006-2007 events) occurred during previously-reported 'global upheavals' of Jupiter's belt/zone structure, where the colours appeared to change in sequence (Rogers, 1995; Rogers, 2007). However, further characterizations of new equatorial disturbances, and the changes occurring within the neighbouring belts, will be required to understand the possible connections between these different regions of Jupiter.

With either time interval, we expect a new EZ disturbance in 2019-21, during the epoch of Juno's exploration, as a strong brown coloration event is currently (spring 2018) developing in the EZ (see Figure 1L and Figure S5 in the online supporting information). New close-in observations of the EZ at visible (JunoCam) and infrared wavelengths (2-5 µm using JIRAM), and with the Juno Microwave Radiometer (MWR) sensing below the clouds will be able to measure possible changes to the deep $NH_3$ plume at the EZ (Li et al., 2017). If such an event does occur as predicted, Juno has the potential to significantly improve our understanding on the origin of EZ disturbances at 5 µm and their relationship to the observed cloud morphology and coloration.

**Acknowledgements**

Data can be accessed from the following GitHub repository http://doi.org/10.5281/zenodo.1412629, which contains the cylindrical maps from 1999-2000 and 2006-2007 and the normalised brightness scans from ±7° latitude as a function of time. This work is partially supported by a European Research Council Consolidator Grant under the European Union's Horizon 2020 research and innovation programme, grant agreement number 723890, at the University of Leicester, by a Royal Society Research Fellowship and by UK Science and Technology Facilities Council (STFC) Grant ST/N000749/1. A portion of this work was performed by GSO at the Jet Propulsion Laboratory, California Institute of Technology, under a contract with NASA. We are grateful to thank all those involved in the acquisition of these 5-μm data over many years, including but not limited to Thomas Greathouse, Kevin Baines, Padma Yanamandra-Fisher, Tom Momary, Jim Friedson, Jose Luis Ortiz and John Spencer. We wish to thank to the following amateur astronomers for their contributions to the visible-light imaging: Christopher Go, Tiziano Olivetti, Isao Miyazaki and Hans-Jorg Mettig. This investigation is based on data acquired at (i) the Infrared Telescope Facility, operated by the University of Hawaii under